\documentclass[iop]{emulateapj}
\usepackage{apjfonts}
\usepackage{amsmath}

\begin{document}
%\linenumbers

\title{Dust formation in macronovae}

\author{Hajime Takami\altaffilmark{1,4}, Takaya Nozawa\altaffilmark{2}, Kunihito Ioka\altaffilmark{1,3}}

\altaffiltext{1}{Institute of Particle and Nuclear Studies, KEK, 1-1, Oho, Tsukuba 305-0801, Japan; e-mail: takami@post.kek.jp, kunihito.ioka@kek.jp}
\altaffiltext{2}{National Astronomical Observatory of Japan, 2-21-1, Osawa, Mitaka, Tokyo 181-8588, Japan; takaya.nozawa@nao.ac.jp}
\altaffiltext{3}{Department of Particle and Nuclear Physics, the Graduate University for Advanced Studies, Tsukuba 305-0801, Japan}
\altaffiltext{4}{JSPS Research Fellow}

\shorttitle{Dust formation in macronovae}
\shortauthors{Takami, Nozawa, \& Ioka}

\begin{abstract}
We examine dust formation in macronovae (as known as kilonovae), which are the bright ejecta of neutron star binary mergers and one of the leading sites of r-process nucleosynthesis. We find that dust grains of r-process elements are difficult to form because of the low number density of the r-process atoms, while carbon or elements lighter than irons can condense into dust if they are abundant, in light of the first macronova candidate associated with GRB 130603B. Dust grains absorb emission from ejecta with opacity even greater than that of the r-process elements, and re-emit photons at infrared wavelengths. Such dust emission can potentially account for the macronova without r-process nucleosynthesis as an alternative model. This dust scenario predicts a more featureless spectrum than the r-process model and day-scale optical-to-ultraviolet emission. 
\end{abstract}

\keywords{binaries: general --- dust, extinction --- gamma-ray burst: individual (GRB 130603B) --- infrared: stars --- methods: numerical --- stars: neutron}

%%%%%%%%%%%%%%%%%%%%%%%%%%%%%%%%%%%%%%%%%%%%%%%%%%%%%%%%%%%%%%%%%%%%%%
%%%%%%%%%%%%%%%%%%%%%%%%%%%%%%%%%%%%%%%%%%%%%%%%%%%%%%%%%%%%%%%%%%%%%%
\section{Introduction} \label{sec:intro}
%%%%%%%%%%%%%%%%%%%%%%%%%%%%%%%%%%%%%%%%%%%%%%%%%%%%%%%%%%%%%%%%%%%%%%
%%%%%%%%%%%%%%%%%%%%%%%%%%%%%%%%%%%%%%%%%%%%%%%%%%%%%%%%%%%%%%%%%%%%%%

Macronovae are brightening phenomena associated with the ejecta from the mergers of neutron star binaries (NSBs), i.e., neutron star (NS)-NS binaries and black hole (BH)-NS binaries. In the original macronova model, the luminosity peaks at $\sim 1$ day after the mergers, with the opacity coefficient of ejecta $\kappa \sim 0.1$ cm$^2$ g$^{-1}$ \citep{Li1998ApJ507L59,Kulkarni2005astro-ph0510256,Metzger2010MNRAS406p2650}. Recent studies have shown that r-process nucleosynthesis occurs efficiently in neutron-rich ejecta \citep[low electron fraction $Y_e$, e.g.,][]{Goriely2011ApJ738L32,Korobkin2012MNRAS426p1940,Bauswein2013ApJ773p78}. The r-process elements, especially lanthanoids, provide large opacity for ejecta \citep[$\kappa \sim 10$ cm$^2$ g$^{-1}$;][]{Barnes2013ApJ775p18,Tanaka2013ApJ775p113,Tanaka2014ApJ780p31}, so that their luminosity peaks around $10$ days (called the r-process model). In both cases, radioactive decay heats the ejecta and powers emission. The r-process model successfully reproduces the near-infrared (NIR) macronova in the afterglow of short gamma-ray burst (GRB) 130603B \citep[$z = 0.356$;][]{Berger2013ApJ774L23,Tanvir2013Nature500p547,Hotokezaka2013ApJ778L16}. NSB mergers are the most promising sources of gravitational waves which are expected to be directly detected by the next generation interferometers, such as advanced LIGO \citep{Abadie2010NIMPhysResSectA624p223}, advanced VIRGO \citep{Accadia2011CQGra28p114002}, and KAGRA \citep{Kuroda2010CQGra27p084004}. The electromagnetic detection of macronovae improves the localization of gravitational wave sources; the localization accuracy by photons is much better than that by the interferometers, $\sim 10$ - $100$ deg$^2$ \citep[e.g.,][]{Aasi2013arXiv1304.0670}.

Recent numerical simulations have revealed that NSB mergers eject significant masses with $\sim 10^{-4} M_{\odot}$ -- $10^{-2} M_{\odot}$ dynamically \citep[e.g.,][]{Rosswog1999A&A341p499,Ruffert2001A&A380p544,Hotokezaka2013PRD87p024001,Bauswein2013ApJ773p78,Kyutoku2013PRD88p041503} and/or by neutrino-driven \citep[e.g.,][]{Ruffert1997A&A319p122,Rosswog2003MNRAS345p1077,Dessart2009ApJ690p1681,Metzger2014arXiv1402.4803} or magnetically driven winds \citep[e.g.,][]{Shibata2011ApJ734L36}, which can explain GRB 130603B in the r-process model. The ejecta also interact with circumstellar matter and radiate like supernova remnants at a later phase \citep{Nakar2011Nature478p82,Piran2013MNRAS430p2121,Takami2013arXiv1307.6805,Kyutoku2014MNRAS437L6}.

Different types of nucleosynthesis may take place in the ejecta, depending on $Y_e$. While r-process nucleosynthesis occurs in low $Y_e$ ejecta, relatively high $Y_e$ ($\sim 0.2$ -- $0.5$) can be also realized, which has been exclusively discussed for neutrino-driven winds \citep[e.g.,][see also \citet{Wanajo2014arXiv1402.7317} for locally low $Y_e$ dynamical ejecta]{Fernandez2013MNRAS435p502,Rosswog2013arXiv1307.2939,Surman2013arXiv1312.1199,Metzger2014arXiv1402.4803}. In such environments, r-process nucleosynthesis is inefficient, but heavy elements (up to $^{56}$Ni) may be synthesized from the constituent nucleons of NS matter, e.g., through a series of captures of $\alpha$ particles by $^{12}$C produced by the triple-$\alpha$ process \citep[e.g.,][]{Surman2013arXiv1312.1199}.

In this Letter, we investigate dust formation in the ejecta of NSB mergers for the first time. The formation of dust in macronovae is expected as in supernovae \citep[e.g.,][]{Nozawa2003ApJ598p785} because heavy elements may be synthesized and the ejecta temperature may be low enough for dust formation. We demonstrate that the newly formed dust can be responsible for the opacity of ejecta and its emission can potentially reproduce the NIR excess of GRB 130603B. Although the r-process model can explain this macronova, it is based on the limited observational data. Thus, it is worth considering the dust scenario \citep[e.g., see ][for another possibility]{Jin2013ApJ775L19}.

%%%%%%%%%%%%%%%%%%%%%%%%%%%%%%%%%%%%%%%%%%%%%%%%%%%%%%%%%%%%%%%%%%%%%%
%%%%%%%%%%%%%%%%%%%%%%%%%%%%%%%%%%%%%%%%%%%%%%%%%%%%%%%%%%%%%%%%%%%%%%
\section{Physical properties of the ejecta} \label{sec:model}
%%%%%%%%%%%%%%%%%%%%%%%%%%%%%%%%%%%%%%%%%%%%%%%%%%%%%%%%%%%%%%%%%%%%%%
%%%%%%%%%%%%%%%%%%%%%%%%%%%%%%%%%%%%%%%%%%%%%%%%%%%%%%%%%%%%%%%%%%%%%%

Dust formation depends on the time evolution of gas density and temperature. We assume uniform ejecta with temperature $T$ and density $\rho$ for simplicity. The NIR macronova of GRB 130603B gives only one observational point \citep{Tanvir2013Nature500p547}: the absolute AB {\it J}-band magnitude of $M(J)_{\rm AB} = -15.35$ at $t \sim 7$ days after GRB 130603B in the rest frame. Ejecta are likely in a free expansion phase with the radius of 
\begin{equation}
R = 3.6 \times 10^{15} \left( \frac{\beta}{0.2} \right) 
\left( \frac{t}{7~{\rm days}} \right)~~~{\rm cm}, 
\end{equation}
where $\beta$ is the ejecta speed divided by light speed $c$, and is basically determined by the escape velocity of NSs. Hereafter, $\beta = 0.2$ is adopted.

We can estimate the effective temperature of the ejecta at $t_0 = 7$ days as $T_0 \sim 2000$ K assuming a blackbody radiation, so that 
\begin{equation}
T = T_0 \left( \frac{t}{7~{\rm days}} \right)^{-s}, 
\label{eq:temp}
\end{equation}
where $s = 1$ for adiabatic expansion, and $s = (\alpha + 2)/4$ for the case with the heating rate $\propto t^{-\alpha}$, which can be derived from the second law of thermodynamics, $T d(T^3 R^3)/dt \propto t^{-\alpha}$ \citep{Li1998ApJ507L59}. Recent detailed studies of r-process nucleosynthesis have suggested $\alpha = 1.2$ - $1.3$ \citep{Metzger2010MNRAS406p2650,Korobkin2012MNRAS426p1940,Rosswog2013arXiv1307.2939,Wanajo2014arXiv1402.7317}. Our results are not sensitive to $s$ because dust-formation temperature is not far from $T_0$.

The density of the ejecta $\rho_0$ at $t_0$ is estimated from the condition that the diffusion time of photons $\sim \rho_0 R^2 \kappa / c$ is comparable with the dynamical time $\sim R / c\beta$ for a given opacity coefficient $\kappa$, which is achieved at around the luminosity peak. The upper limit of the NIR flux at $\sim 20$ days in the rest frame suggests that the luminosity peaks at $\sim 10$ days \citep{Tanvir2013Nature500p547}. The density scales as $\rho = \rho_0 (t / t_0)^{-3}$ and therefore, 
\begin{equation}
\rho \sim 1.4 \times 10^{-16} \left( \frac{\kappa}{10~{\rm cm}^2 {\rm g}^{-1}} 
\right)^{-1} \left( \frac{\beta}{0.2} \right)^{-2} 
\left( \frac{t}{7~{\rm days}} \right)^{-3} ~~~{\rm g~cm}^{-3}. 
\label{eq:rho}
\end{equation}
We examine the two cases for ejecta density. One is low-density ejecta with $\kappa = 10$ cm$^2$ g$^{-1}$, which is the case that r-process elements are efficiently synthesized and are responsible for the opacity of the gas \citep{Kasen2013ApJ774p25,Tanaka2013ApJ775p113}. The other is high-density ejecta with $\kappa = 0.1$ cm$^2$ g$^{-1}$, which corresponds to the opacity coefficient of Fe-rich Type Ia supernovae \citep[e.g.,][]{Pinto2000ApJ530p757}, i.e., inefficient r-process nucleosynthesis.

The corresponding mass of the ejecta $M = \Delta \Omega R^3 \rho / 3$ is 
\begin{equation}
M \sim 1.4 \times 10^{-2} M_{\odot} 
\left( \frac{\Delta \Omega}{4 \pi} \right) 
\left( \frac{\kappa}{10~{\rm cm}^2 {\rm g}^{-1}} \right)^{-1} 
\left( \frac{\beta}{0.2} \right), 
\label{eq:tmass}
\end{equation}
where $\Delta \Omega$ is the solid angle within which the ejecta are blown off. Regarding dynamical ejecta, general relativistic hydrodynamical simulations have indicated $\Delta \Omega / 4 \pi \sim 1$ for NS-NS mergers \citep{Hotokezaka2013PRD87p024001}, and $\Delta \Omega / 4 \pi \sim 0.1$ for BH-NS mergers \citep{Kyutoku2013PRD88p041503}.

%%%%%%%%%%%%%%%%%%%%%%%%%%%%%%%%%%%%%%%%%%%%%%%%%%%%%%%%%%%%%%%%%%%%%%
%%%%%%%%%%%%%%%%%%%%%%%%%%%%%%%%%%%%%%%%%%%%%%%%%%%%%%%%%%%%%%%%%%%%%%
\section{Dust Formation} \label{sec:dform}
%%%%%%%%%%%%%%%%%%%%%%%%%%%%%%%%%%%%%%%%%%%%%%%%%%%%%%%%%%%%%%%%%%%%%%
%%%%%%%%%%%%%%%%%%%%%%%%%%%%%%%%%%%%%%%%%%%%%%%%%%%%%%%%%%%%%%%%%%%%%%

%%%%%%%%%%%%%%%%%%%%%%%%%%%%%%%%%%%%%%%%%%%%%%%%%%%%%%%%%%%%%%%%%%%%%%
\subsection{Dust formation in the ejecta}
%%%%%%%%%%%%%%%%%%%%%%%%%%%%%%%%%%%%%%%%%%%%%%%%%%%%%%%%%%%%%%%%%%%%%%

Dust can form below the equilibrium temperature at which the partial pressure of an element $i$ equals to the vapor pressure of the condensate which depends only on temperature and atomic species. Partial gas pressure can be calculated from $T$ and $\rho_i = f_i \rho$ with the equation of state of ideal gas, where $f_i$ is the mass fraction of the element $i$. This study focuses on representative elements in Table \ref{tab:thermodata} individually instead of complicated composition to avoid complexity. Sr and Pt are elements in the first and third peaks of the r-process nucleosynthesis, respectively. Hf is taken as a substitution of lanthanoid elements\footnote{The data necessary for dust formation calculations are not available for lanthanoid elements. The second peak element of r-process is Xe, but the formation of Xe dust is difficult because it is a noble gas.}.

One of the conditions necessary for dust formation is that the timescale of collisions between atoms $\tau_{\rm coll} = [ \pi a_0^2 n_i \langle v \rangle ]^{-1}$ is much shorter than that of the expansion of the ejecta $\tau_{\rm exp} = | (1/\rho)(d\rho/dt) |^{-1}$. Here, $a_0 \sim 1~\mathring{\rm A}$, $n_i = \rho_i / m_i$, $m_i = A_i m_{\rm H}$, and $A_i$ are the radius, number density, mass, and molecular weight of the element $i$, respectively. $m_{\rm H}$ is the mass of a hydrogen atom. The mean thermal velocity is represented by $\langle v \rangle = (2 k_{\rm B} T / m_i)^{1/2}$ with the Boltzmann constant $k_{\rm B}$. Then, using equations (\ref{eq:temp}) and (\ref{eq:rho}), the ratio $\tau_{\rm coll}/\tau_{\rm exp}$is 
\begin{eqnarray}
\frac{\tau_{\rm coll}}{\tau_{\rm exp}} &=& 
\frac{3 A_i m_{\rm H}}{\pi a_0^2 f_i \rho_0 t_0} 
\left( \frac{A_i m_{\rm H}}{2 k_{\rm B} T_0} \right)^{1/2} 
\left( \frac{T}{T_0} \right)^{-\frac{2}{s} - \frac{1}{2}} \nonumber \\
%&\simeq& 0.32 f_{\rm i}^{-1}
&\simeq& 0.45 f_{\rm i}^{-1}
\left( \frac{A_i}{100} \right)^{\frac{3}{2}} 
%\left( \frac{\rho_0}{1.4 \times 10^{-16}~{\rm g~cm}^{-3}} \right)^{-1} \nonumber \\
\left( \frac{\rho_0}{10^{-16}~{\rm g~cm}^{-3}} \right)^{-1} \nonumber \\
&& ~~~~~~~~~~~~~~~~~~~~~ \times 
\left( \frac{T}{2000~{\rm K}} \right)^{-\frac{2}{s} - \frac{1}{2}}. 
\label{eq:ness}
\end{eqnarray}
This ratio is sensitive to the temperature ($\propto T^{-3.0}$ for $\alpha = 1.2$), and increases with time. A dust grain does not form if this ratio is above unity. For instance, the ratio is $\sim 1$ in the low density ejecta with $f_i \sim 0.5$ and $A_i \sim 200$, which indicates that dust grains of r-process elements are difficult to form.

%%%%%%%%%%%%%%%%%%%%%%%%%%%%%%%%%%%%%%%%%%%%%%%%%%%%%%%%%%%%%%%%%%%%%
%\begin{deluxetable}{lcccc}
\begin{deluxetable}{lccccc}
%\tabletypesize{\scriptsize}
\tablewidth{0pt}
\tablecaption{Grain Species Considered in the Calculations}
\tablehead{ 
\colhead{Grain Species} & \colhead{$\gamma/10^4$K} 
& \colhead{$\delta$} & \colhead{$a_0$ ($\mathring{A}$)} 
& \colhead{$\sigma$ (erg cm$^{-2}$)}
%}
& \colhead{$T_{\rm eq}$ (K)}
}
\startdata
\\
C   & ~8.640 & 18.974 & 1.281 & 1400 & 1800--2000 \\
Fe  & ~4.842 & 16.557 & 1.411 & 1800 & 1060--1180 \\
Sr  & ~1.456 & ~7.067 & 2.364 & ~165 & 430--500 \\
%
%Zr  & ~7.294 & 16.579 & 1.770 & 2080 & 1600--1780 \\
%
%Ba  & ~1.802 & ~8.146 & 2.494 & ~195 & 500--560 \\
%
%W   & 10.243 & 17.085 & 1.586 & 3200 & 2230--2490 \\
%
Hf  & ~7.390 & 16.220 & 1.745 & 1510 & 1630--1820 \\
Pt  & ~6.777 & 17.826 & 1.533 & 1770 & 1440--1600 \\
%
%Pb  & ~2.040 & ~9.329 & 1.935 & ~410 & 540--610 \\
%
\enddata
\tablecomments{
The Gibbs free energy for the formation of the condensate is approximated by ${\it \Delta} \mathring{g} / k T = - \gamma/T + \delta$, where the numerical values $\gamma$ and $\delta$ are derived by least-squares fittings of the thermodynamics data \citep{Chase1985JPhysChemRefDataSupp14p1,Arblaster2005PlatinumMetalsRev49p141} in the temperature range of $T =$ 200--2500 K. The radius of the condensate per atom and the surface tension of the condensate are given as $a_0$ and $\sigma$, respectively \citep{Elliott1980book,Nozawa2003ApJ598p785}. The equilibrium temperatures $T_{\rm eq}$ are presented for the density range of $n_i = 10^6$--$10^8$ cm$^{-3}$. See \citet{Nozawa2013ApJ776p24} for the details of the dust formation calculations.}
\label{tab:thermodata}
\end{deluxetable}
%%%%%%%%%%%%%%%%%%%%%%%%%%%%%%%%%%%%%%%%%%%%%%%%%%%%%%%%%%%%%%%%%%%%%

\begin{figure}
\includegraphics[width=0.95\linewidth]{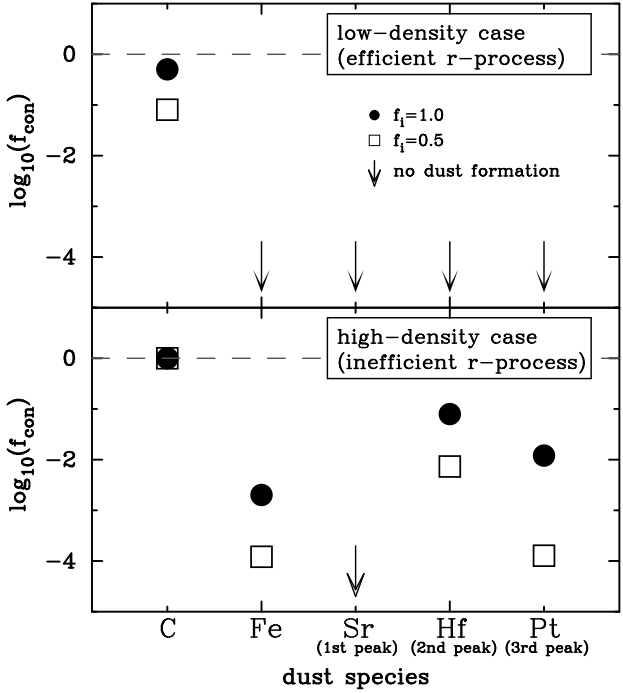}
\caption{Condensation efficiency, defined as the mass fraction of atoms finally locked up in dust grains, of carbon, iron, strontium, hafnium, and platinum in the low-density ({\it upper}) and high-density ({\it lower}) cases. Note that the dust formation of Hf and Pt in the high-density case is unlikely because low $f_i$ ($\lesssim 10^{-2}$) is required for r-process elements in GRB 130603B.}
\label{fig:dfcon}
\end{figure}

We calculate dust formation by applying the formulation of non-steady-state dust formation in \citet{Nozawa2013ApJ776p24} with $s = 0.8$, i.e., $\alpha = 1.2$ under $T$ and $\rho$ given in Section \ref{sec:model}. The results are summarized in Figure \ref{fig:dfcon}. In the low-density ejecta Sr, Hf, and Pt never condense into dust grains even for $f_i = 1.0$. Here, dust grains are defined as clusters containing atoms more than a hundred. Fe grains also cannot be formed even if $f_i = 1.0$ because of its low condensation temperature ($\la 1000$ K). Note that condensation temperature is lower than the equilibrium temperature because supersaturation is required for dust formation. The formation of carbon grains is possible as a result of the high condensation temperature ($\ga$ 1800 K) and small molecular weight (see Equation \ref{eq:ness}), i.e., large number density, which achieves $f_{\rm con} = 0.08$ and $0.5$ for $f_i = 0.5$ and $1.0$, respectively. These carbon grains could be responsible for the opacity of ejecta if a significant amount of carbon exists. We conclude that the dust grains of heavy ($A_i \gtrsim 50$) elements are not expected to be formed in the low-density case. In particular, the condensation of r-process elements ($A_i \gtrsim 100$) is extremely difficult, unless the inhomogeneity of ejecta is taken into account.

In the high-density ejecta (see Equation \ref{eq:rho}), the necessary condition for dust formation is relaxed. Carbon grains form at $t \sim 7$ days with $f_{\rm con} = 1.0$ for $f_i = 0.5$. Despite the high density, the low condensation temperature permits Fe grains only below $f_{\rm con} = 2 \times 10^{-3}$ and no Sr grains even for $f_i = 1.0$. Dust grains of Hf and Pt could form if $f_i$ is high enough. However, for GRB 130603B, such a high $f_i$ is unlikely because a significant amount of r-process elements leads to $\kappa$ much larger than $0.1$ cm$^2$ g$^{-1}$, which is far from the definition of the high-density case ($\kappa \sim 0.1$ cm$^2$ g$^{-1}$) and therefore is inconsistent with the observation. Since Fe grains form much later than $t_0 = 7$ days\footnote{Fe grains could form at earlier time in non-uniform ejecta, e.g., at the head of the ejecta where gas is rapidly cooled.}, we exclusively focus on carbon grains for GRB 130603B in the next subsection.

%%%%%%%%%%%%%%%%%%%%%%%%%%%%%%%%%%%%%%%%%%%%%%%%%%%%%%%%%%%%%%%%%%%%%%
\subsection{Dust model for macronovae}
%%%%%%%%%%%%%%%%%%%%%%%%%%%%%%%%%%%%%%%%%%%%%%%%%%%%%%%%%%%%%%%%%%%%%%

\begin{figure}
\includegraphics[width=0.95\linewidth]{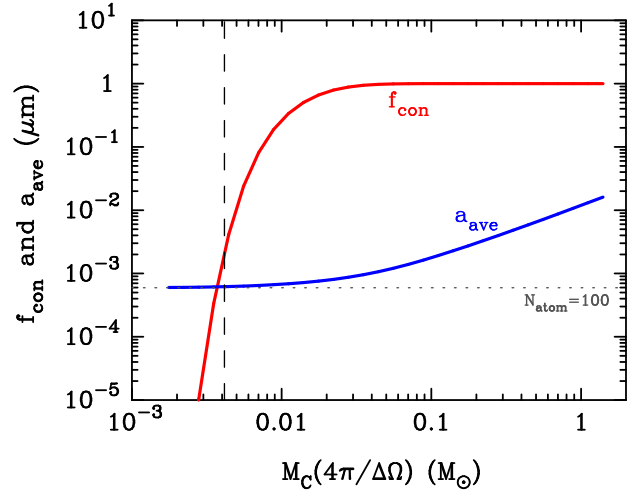}
\caption{Condensation efficiency and average radius of newly formed carbon grains as a function of the isotropic mass of carbon $M_{\rm C} (4\pi / \Delta \Omega)$ ($M_{\rm C}$ is the carbon mass available for dust formation). The radius of carbon grains with 100 atoms is also shown ({\it dotted}). The vertical dashed line marks the amount of carbon gas which is required to produce the dust emission spectrum ($M_{\rm d,C} = 8 \times 10^{-6} M_{\odot}$) in Figure \ref{fig:spec} for $\Delta \Omega = 4\pi$.}
\label{fig:cabun}
\end{figure}

Figure \ref{fig:cabun} shows the condensation efficiency and average radius of carbon grains in the high-density ejecta (see Equation \ref{eq:rho}) as a function of the isotropic equivalent mass of carbon $M_{\rm C} (4\pi / \Delta \Omega)$, where $M_{\rm C}$ is the mass of carbon available for dust formation. The density of ejecta is calculated as the isotropic equivalent mass divided by $4\pi R^3 / 3$. The condensation efficiency is unity above $M_{\rm C} (4\pi / \Delta \Omega) \sim 10^{-2}$, below which it decreases rapidly.

\begin{figure}
\includegraphics[width=0.95\linewidth]{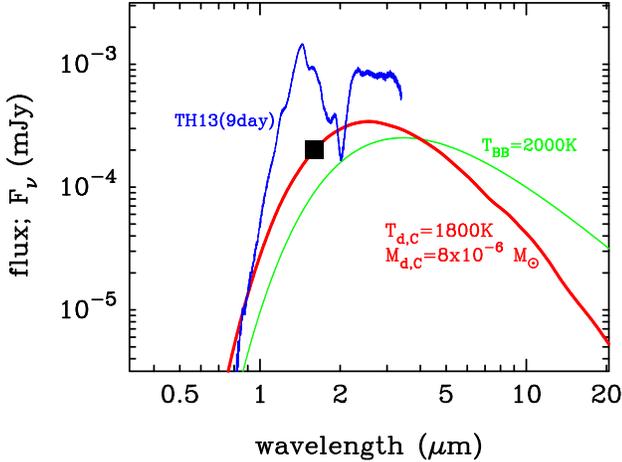}
\caption{Thermal emission spectrum of carbon grains with $1800$ K ({\it red}) in the observer frame. The black square is the observed NIR flux of GRB 130603B \citep{Tanvir2013Nature500p547}. For references, a spectrum of the r-process model \citep[{\it blue};][TH13]{Tanaka2013ApJ775p113} and a blackbody spectrum with $T = 2000$ K ({\it green}) are shown. The spectral shape of the dust emission is rather smooth than that of the r-process model. Note that the parameters of TH13 are not optimized for GRB 130603B.}
\label{fig:spec}
\end{figure}

Once dust grains are formed in the ejecta, they absorb emission from the ejecta and emit photons with energies corresponding to their temperature. Assuming that dust emission is optically thin, we can choose the mass and temperature of dust so that its thermal emission explains the NIR emission of GRB 130603B. Figure \ref{fig:spec} demonstrates one of the examples of the dust emission spectrum. The required mass of carbon grains is $M_{\rm d,C} \sim 8 \times 10^{-6} M_{\odot}$ with the dust temperature of 1800 K. Since $M_{\rm d,C} = f_{\rm con} M_{\rm C}$, the condensation efficiency for achieving this dust mass is estimated as $f_{\rm con} = 2 \times 10^{-3}$ ($M_{\rm C} = 4 \times 10^{-3} M_{\odot}$; see Figure \ref{fig:cabun}) for isotropic ejecta, i.e., $\Delta \Omega = 4\pi$. For anisotropic ejecta ($\Delta \Omega < 4\pi$), a specific $M_{\rm C}$ leads to $f_{\rm con}$ larger in the cases of smaller $\Delta \Omega$ because of the higher ejecta density. Thus, a smaller amount of $M_{\rm C}$ is sufficient to obtain a fixed value of $M_{\rm d,C}$ for a smaller $\Delta \Omega$. We can estimate $f_{\rm con}$ and $M_{\rm C}$ to yield a given $M_{\rm d,C} = f_{\rm con} M_{\rm C}$ by using the red line in Figure \ref{fig:cabun}. For instance, in the case of $( 4\pi / \Delta \Omega ) = 10$, $f_{\rm con} \sim 2 \times 10^{-2}$, which is achieved when the mass of carbon gas is $M_{\rm C} \sim 5 \times 10^{-4} M_{\odot}$. Note that the required mass of carbon gas is consistent with the mass of ejecta which is shown in recent simulations. We also notice that carbon grains do not evaporate because the dust temperature is lower than its evaporation (equilibrium) temperature.

The opacity coefficient of the carbon grains is $\kappa_{\rm C} \sim 1.1 \times 10^4$ cm$^{2}$ g$^{-1}$ at the {\it J}-band ($1.2 \mu$m), which does not depend on their radius as long as the grain radius is much smaller than the wavelength. At 7 days in the rest frame, the absorption probability is above unity at wavelengths shorter than the {\it J}-band under the parameter choice : $( 3 M_{\rm d,C} / \Delta \Omega R^3) \kappa_{\rm C} R \beta \sim 0.7 (4\pi / \Delta \Omega) (\lambda / 1.2~\mu{\rm m})^{-1.2}$. The absorption (emission) coefficients are proportional to $\lambda^{-1.2}$ \citep{Zubko1996MNRAS282p1321}. Thus, the dust grains absorb macronova emission and re-emit radiation at NIR wavelengths.

Compared to the r-process model, which predicts a rather broad-line spectrum, the dust model provides an even more featureless spectrum. The spectrum of dust emission also deviates from a blackbody spectrum at wavelengths of $\gtrsim 5 \mu$m because of the $\lambda$-dependence of the emission coefficient. Thus, a spectrum of the dust model can be observationally distinguished from both the r-process model and a pure blackbody spectrum.

\begin{figure}
\includegraphics[width=0.95\linewidth]{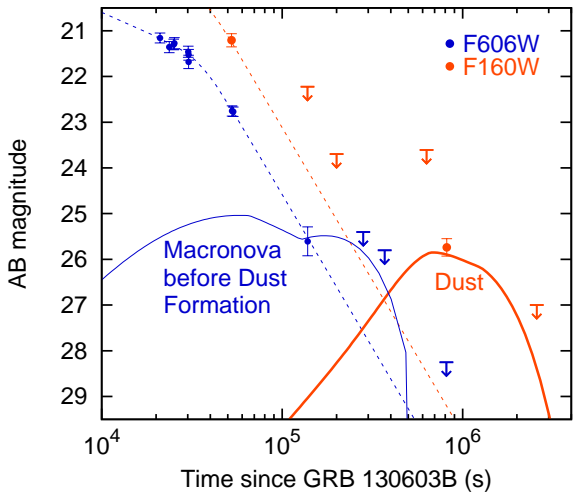}
\caption{Schematic light curves for the interpretation of GRB 130603B by the dust model. The data points and fits by a GRB afterglow model ({\it dashed lines}) are from \citet{Tanvir2013Nature500p547}.} 
\label{fig:lc}
\end{figure}

Based on the discussion above, we can propose an alternative model for the NIR brightening of GRB 130603B, and generally macronovae, which is schematically drawn in Figure \ref{fig:lc}. First, a macronova that is not powered by r-process elements happens. The low line opacity coefficient, e.g., $\kappa \sim 0.1$ cm$^2$ s$^{-1}$, makes this event bright in a blue band on timescale of a few day, as in the original work \citep{Li1998ApJ507L59}. Then, when the temperature of ejecta decreases to $\sim 2000$ K, dust formation begins and the ejecta become opaque again by dust. The high opacity allows the macronova emission to heat dust grains, which emit absorbed photons at NIR wavelengths. In other words, the observed NIR macronova can be the dust emission. The dust model has a featureless spectrum and an early macronova in a blue band, compared to the r-process model with the broad-line spectrum and without early blue emission.

%%%%%%%%%%%%%%%%%%%%%%%%%%%%%%%%%%%%%%%%%%%%%%%%%%%%%%%%%%%%%%%%%%%%%%
%%%%%%%%%%%%%%%%%%%%%%%%%%%%%%%%%%%%%%%%%%%%%%%%%%%%%%%%%%%%%%%%%%%%%%
\section{Discussion and Summary} \label{sec:dissum}
%%%%%%%%%%%%%%%%%%%%%%%%%%%%%%%%%%%%%%%%%%%%%%%%%%%%%%%%%%%%%%%%%%%%%%
%%%%%%%%%%%%%%%%%%%%%%%%%%%%%%%%%%%%%%%%%%%%%%%%%%%%%%%%%%%%%%%%%%%%%%

We have investigated dust formation in macronovae based on the temperature and density estimated from GRB 130603B. We have shown that dust of r-process elements hardly form even if they are abundantly produced. On the other hand, dust of light elements such as carbon can be formed. We have also suggested that the NIR macronova of GRB 130603B can be explained by the emission of light-element dust such as carbon grains, as an alternative to the r-process model.

We inferred the temperature of ejecta from the observational result as $T_0 \sim 2000$ K. A heating source may be the radioactive decay products of r-process elements in the r-process model. In r-process nucleosynthesis inefficient ejecta one possibility is the radioactive decay of heavy, but not r-process, elements \citep{Barnes2013ApJ775p18}. Radioactive nuclei with the lifetime less than $\sim 10$ days release a significant fraction of radioactive energies and achieve $T \sim 2000$ K at $\sim 7$ days under a reasonable choice of ejecta mass. Shock heating may be also possible.

We should keep in mind that the discussions in this Letter are based on the observational result of GRB 130603B because this is the only existing sample. For instance, if ejecta temperature is lower and density is much higher in another macronova, dust grains of r-process elements could be formed.

Our results have shown that newly formed grains are relatively small, consisting of $\sim 100$ up to $\sim 10^5$ atoms. We adopted the theory of Mie scattering in calculating the absorption coefficients of dust. However, they might deviate from the prediction of the theory for the dust only containing order-of-hundreds atoms.

We have considered homogeneous ejecta for simplicity. In reality, ejecta may be inhomogeneous, and dust may be formed in dense clumps, as discussed in supernovae \citep[e.g.,][]{Kotak2009ApJ704p306,Indebetouw2014ApJ782L2}. Larger dust grains may be formed in higher density clumps, and then opacity by dust can be changed by reflecting the spatial distribution of dust. Moreover, the consideration of the radial profile of ejecta may modify a dust formation history. Such effects are interesting subjects to be studied in the future.

The dust model for NIR macronovae should be tested observationally. One is the confirmation of a featureless spectrum. As shown in Figure \ref{fig:spec}, a dust emission spectrum is even featureless compared to a broad spectrum in the r-process model. A spectrum in the dust model also deviates from a blackbody spectrum at long wavelengths. Another way is multi-wavelength observations of light curves from early epochs (see Figure \ref{fig:lc}). Without opacity of r-process elements, a macronova is bright and blue in an early phase. It becomes red later by the emission of newly formed dust. Early optical emission was explored for a few short GRBs \citep[e.g., GRB 050509B;][]{Hjorth2005ApJ630L117}. Although the flux limit is strong for these GRBs, continuous searches for early emission are important because the properties of ejecta in NSB mergers may not be universal, e.g., depending on progenitors (NS-NS / BH-NS). In both cases, quick follow-up observations are important to understand the origin of NIR macronovae as well as the nucleosynthesis in the mergers of compact stellar objects.

\acknowledgements 
We thank J.~Hjorth and N.~R.~Tanvir for useful comments, M.~Tanaka for providing us with a macronova spectrum (Figure \ref{fig:spec}), and K. Hotokezaka, T. Nakamura, Y.~Sekiguchi, and M. Shibata for discussions. T. Nakamura had also investigated dust formation in macronova independently of us. This work is supported by KAKENHI $24 \cdot 9375$ (H.~T.), $22684004$, $23224004$, $26400223$ (T.~N.), $24000004$, $22244030$, $24103006$, and $26287051$ (K.~I.), and by World Premier International Research Center Initiative (WPI Initiative), MEXT, Japan.

\bibliographystyle{apj}
%\bibliography{ms.bib}

%\end{document}

\end{document}